\documentclass[pra,twocolumn,superscriptaddress,10pt,noshowpacs]{revtex4}
\usepackage[english]{babel}
\usepackage[T1]{fontenc}
\usepackage[utf8]{inputenc}
\usepackage{graphicx,epstopdf}
\usepackage{amssymb}
\usepackage{amsmath}

\usepackage{amsfonts}
\usepackage{bbm}
\usepackage{color}
\usepackage{latexsym}
\usepackage{caption}
\usepackage{subcaption}
\usepackage{times,txfonts}
\usepackage{color,soul}
\usepackage{listings}
\usepackage{bbold}

\newcommand{\beq}{\begin{equation}}
\newcommand{\eeq}{\end{equation}}
\newcommand{\bea}{\begin{eqnarray}}
\newcommand{\eea}{\end{eqnarray}}
\begin{document}

\title{Black string solutions in Lifshitz spacetime}

\author{L. A. Lessa}
\email{leandrolessa@fisica.ufc.br}
\affiliation{Universidade Federal do Cear\'a, Departamento de F\'{i}sica, 60455-760, Fortaleza, CE, Brazil}

\author{J. E. G. Silva}
\email{euclides@fisica.ufc.br}
\affiliation{Universidade Federal do Cear\'a, Departamento de F\'{i}sica, 60455-760, Fortaleza, CE, Brazil}

\author{J. Furtado}
\email{job.furtado@ufca.edu.br}
\affiliation{Universidade Federal do Cariri, Centro de Ci\^encias e Tecnologia, 63048-080, Juazeiro do Norte, CE, Brazil}
\affiliation{Department of Physics, Faculty of Science, Gazi University, 06500 Ankara, Turkey}

\begin{abstract}

In this paper we study black string solutions considering the Lifshitz anisotropic scaling.  We have shown that a new class of asymptotically Lifshitz solutions can be generated by an Einstein-Maxwell-Dilaton theory with a cosmological constant. In the limit where we recover conformal scale invariance, we retrieve the usual black string solution. Furthermore, we demonstrated that to incorporate the effects of electric charge in the black string, at least two independent gauge fields coupled to the dilaton field are necessary. The charged black string solution exhibits new horizons that depend on the potential in Lifshitz exponent $z$. The stability of these new solutions is investigated through the thermodynamic analysis of the charged black string. The temperature, entropy, and heat capacity indicate that these modified black strings are thermodynamically stable.
\end{abstract}

\maketitle


\section{Introduction}
Gauge/gravity duality studies have been proving to be a very promising area in theoretical physics in recent years. 
This is mainly due to the operational ease that such a tool offers to obtain weakly coupled and computable dual descriptions of strongly coupled conformal theories \cite{maldacena1999large}. The strongly-coupled systems exhibits a scaling symmetry near critical points. When the critical fixed point is not dynamic, the more familiar scale invariance which arises
in the conformal group is given by
\begin{equation}\label{tr1}
    t \rightarrow \lambda t, \ \      x_i \rightarrow \lambda x_i
\end{equation}
where $\lambda$ is a real constant, the $t$ is time coordinate and $x_i$ are spatial coordinates. In the context of asymptotically anti-de Sitter (AdS) spacetimes that satisfy the invariant property described by (\ref{tr1}), the AdS/CFT correspondence proves to be a valuable tool. This correspondence highlights the significance of such invariances in theoretical physics. Over time, the AdS/CFT correspondence has gained substantial traction, as evidenced by numerous studies applying this duality to various physical scenarios \cite{co1,co2,co3,co4}.

Our primary interest, however, lies in the development of gravitational dual descriptions for models that exhibit anisotropic scale invariancei \cite{kachru2008gravity}, i.e.,
\begin{equation}\label{tr2}
      t \rightarrow \lambda^z t, \ \      x_i \rightarrow \lambda x_i \ \  r \rightarrow \lambda^{-1}r,
\end{equation}
where $z$ is called the Lifshitz exponent. The aforementioned scale symmetry is referred to as the Lifshitz scale symmetry \cite{balasubramanian2008gravity}. For $z=1$, the scaling is isotropic, which corresponds to relativistic invariance.   It has been proposed in \cite{kachru2008gravity,Mann:2009yx} the gravity duals of field theories with Lifshitz scaling should possess metric solutions that exhibit asymptotic behavior of the following form:
\begin{equation}\label{metric}
   ds^2 = - \frac{r^{2z}}{l^{2z}}  dt^2 + \frac{l^2dr^2}{r^2} + r^2 d\Omega^2_{d-1} .
\end{equation}
where $d\Omega^2_{d-1}$ the metric of a unit-radius $S^{d-1}$. 

The proposal presented herein entails a metric characterized by cylindrical and static symmetry with Lifshitz scaling. So that the metric ansatz suitable for this purpose is formulated as follows:
   \begin{equation}\label{metric1}
   ds^2 = - \frac{r^{2z}}{l^{2z}} dt^2 + \frac{l^2dr^2}{r^2} +r^2d\phi^2 + \frac{r^2}{l^2} dy^2.
\end{equation} 
where $t\in (-\infty, \infty)$, the radial coordinate $r\in (0, \infty)$, the axial coordinate $y\in (-\infty, \infty)$ and angular coordinate $\phi \in [0, 2\pi)$. Note that the metric above is invariant under transformations (\ref{tr2}). Additionally, we have three Killing vectors $\frac{\partial}{\partial t }, \frac{\partial}{\partial \phi}, \frac{\partial}{\partial y }$, indicating that indeed we have a space-time with cylindrical-static symmetry \cite{Stephani:2003tm}. In general, the black string solutions of Einstein gravity have been extensively analyzed in the many literatures, see the Refs. \cite{Horne:1991gn,Lemos:1994xp,Darlla:2023qgf}. However, the literature still lacks a black string solution that asymptotically approaches the Lifshitz solution with cylindrical symmetry (\ref{metric1}).

The objective of this work is to generate a black string solution that accounts for the Lifshitz anisotropic scaling (\ref{tr2}). As we will demonstrate later, this exact solution can be derived within the framework of Einstein-Maxwell-dilaton theories (EMDT). These theories are well-suited for generating asymptotically Lifshitz solutions, as Lifshitz spacetime cannot be supported by vacuum solutions and requires matter fields for its generation. We find black hole and black brane solutions in Lifshitz spacetime in EMDT in Refs. 
\cite{Lim:2019hci,Tarrio:2011de}. In Ref. \cite{Lessa:2023dbd}, Lifshitz black holes with Lorentz violation are presented. Significant solutions incorporating dilation, but lacking Lifshitz symmetry, are discussed in Refs. \cite{d1,d2}. Moreover, the dilaton field can naturally arise through the Kaluza-Klein (KK) reduction over the extra dimensions in higher-dimensional theories, see Ref.\cite{Lessa:2023yvw}.

This paper is organised as follows. In section \ref{2}, we obtain an asymptotically Lifshitz black string solution, as well as its charged version. Subsequently, we analyze some geometric properties of the solution. In section \ref{3}, we study the thermodynamics of the solution found and then we analyze its stability. The paper concludes in section \ref{con}. We shall also take the Lorentzian signature for the spacetime metric to be $(-, +, +, +)$.


\section{black string geometry in Lifshitz gravity} \label{2}
In this section, our aim is to seek for a (3+1)-dimensional black string solutions that exhibit asymptotically Lifshitz behavior. 

Let us consider a static and cylindrically symmetric line element in the form
  \begin{equation}\label{metric11}
   ds^2 = - \frac{r^{2z}}{l^{2z}} f(r) dt^2 + \frac{l^2dr^2}{r^2 f(r)} +r^2d\varphi^2 + \frac{r^2}{l^2} dy^2.
\end{equation} 
where the $f(r)$ is the blackening function. In addition, we impose the condition that the function $f(r)$ satisfies the following requirement:
\begin{equation}\label{assin}
\lim_{r \rightarrow \infty} f(r) = 1,
\end{equation}
ensuring that the metric asymptotically approaches the form given by (\ref{metric1}). 

The action considered to govern the black string dynamics is given by an Einstein-dilaton-Mawxell model in the form \cite{Lim:2019hci,Tarrio:2011de}
 \begin{align} \label{action1} 
     S = \frac{1}{16\pi G}\int d^{4} x \sqrt{-g}\bigg[R - \frac{1}{2}(\partial\phi)^2 -2 \Lambda_0 e^{-2\xi_0\phi} -\frac{1}{4}F^2 e^{-2\xi_1\phi}\bigg].
 \end{align}
where $G$ is the Newton gravitational constant and $\Lambda_0$ is the cosmological constant. The first term of the above action is the Einstein-Hilbert term. The scalar field $\phi$ is called the dilaton field that couples to the matter fields of the theory, where $\xi$ are the coupling constants that measure this interaction. Further, the strength field of gauge field is defined as $F_{\mu\nu}=\partial_{[\mu}A_{\nu]}$.

Through the of action (\ref{action1}), we can derive the equations of motion by varying the action with respect to the metric and the fields. This leads us to the following equations, respectively:
\begin{align}\label{eq1} 
    R_{\mu\nu} = \Lambda_0 e^{-2\xi_0\phi}g_{\mu\nu} + \frac{e^{-2\xi_1\phi}}{2}T^{EM}_{\mu\nu} + \frac{1}{2}\partial_{\mu}\phi \partial_{\nu}\phi,
\end{align}
\begin{equation}\label{eq3}
    D_{\mu}(e^{-2\xi_1 \phi}F^{\mu\nu})=0,
\end{equation}
\begin{align}\label{eq4}
  & \Box \phi + 4 \xi_0 \Lambda_0 e^{-2\xi_0\phi} + \frac{\xi_1}{2}F^2 e^{-2\xi_1\phi} =0.
\end{align}
where $\Box=g^{\mu\nu}D_{\nu}D_{\mu}$ and 
\begin{equation}\label{eq6}
    T^{EM}_{\mu\nu} =F_{\mu} \ ^{\sigma}F_{\nu\sigma}-\frac{1}{4}g_{\mu\nu}F^2.
\end{equation}
Note that we used in Eq. (\ref{eq1}) the fact that the gauge field above have traceless in 4-dimensions.

Let us assume an electric field configuration, i.e., $F_{rt}\neq 0$. From the Maxwell equation (\ref{eq3}), in the ansatz metric (\ref{metric11}), we obtain:
\begin{equation}\label{camp2}
F_{rt} = \frac{q e^{2\xi_1 \phi}}{r^{3-z}},
\end{equation}
where  $q$ is a integration constant. The constant $q$ is related to the total charge through further consideration \cite{Tarrio:2011de}
\begin{equation}\label{carga2}
    Q = \frac{1}{16\pi G} \int e^{-2\xi_1\phi}\star F.
\end{equation}
where $\star$ is Hodge star operator.

Note that we have a set of 5 differential equations (\ref{eq1}),(\ref{eq3}) and (\ref{eq4}) and 3 functions $f, \phi$, and $F_{rt}$ to be determined, in addition to 4 parameters $\xi_{1,2}$, $q$, and $\Lambda$ to be fixed. In this regard, we can observe that the system is potentially integrable. Initially, we adopt the standard procedure, as outlined in some instances \cite{Chan:1995fr,d1,Cai:1997ii}, to determine the dilaton field. Thus, we shall substitute the solution (\ref{camp2}) in the components $tt$ and $rr$ of Eq.(\ref{eq1}), we can find from the combination $R^{t}\ _{t}-R^{r}\ _{r}$ that
\begin{equation} \label{dilaton}
    \phi = \phi_0 + \phi_1 \text{ln} r,
\end{equation}
where $\phi_0 $ is a integration constant and $\phi_1$ is given by
\begin{equation}
    \phi_1 =  2\sqrt{z-1}.
\end{equation}
From the expression for $\phi_1$ presented above, it is evident that we require $z\geq1$. Once the expression for the scalar field is obtained, we can substitute it into one of the equations of the components $tt$ and $rr$ of Eq.(\ref{eq1}), then utilize it to eliminate the second derivative of the function $f$ in the component $\varphi \varphi$ or $yy$ of Eq.(\ref{eq1}). Consequently, we are left with solving only a first-order equation for $f$. In conclusion, we thus find that
\begin{eqnarray}\label{ff}
    \nonumber f(r) &=& -\frac{m}{r^{z+2}} +\frac{q^2 e^{2 \xi_1 \phi_0} (\xi_1 \phi_1-1) l^{2 z} r^{2 \xi_1 \phi_1-4}}{4 z (2 \xi_1 \phi_1+z-2)}\\
    &&-\frac{\Lambda_0  l^2 e^{-2 \xi_0 \phi_0} (\xi_0 \phi_1+1) r^{-2 \xi_0 \phi_1}}{z (-2 \xi_0 \phi_1+z+2)}
\end{eqnarray}
where $m$ is a integration constant. To ensure that the limit (\ref{assin}) is obeyed, we adopt
\begin{equation}\label{xo}
    \xi_0=0.
\end{equation}
Therefore, we need to fix the cosmological constant as
\begin{equation}\label{cos}
    \Lambda_0=-\frac{(z+1) (z+2)}{2 l^2}.
\end{equation}
In this case, as we will see later, we can recover a solution that is indeed asymptotically Lifshitz. Also note that for this choice, we find that the dilatonic field decouples from the cosmological constant. After doing so, we need to satisfy Eq. (\ref{eq4}). Substituting (\ref{ff}), (\ref{cos}), and (\ref{xo}) into Eq. (\ref{eq4}), we obtain the following algebraic equation
\begin{equation}\label{eq}
    4(1+z)(2+z)\phi_1 - e^{2\xi_1\phi_0}l^{2z}q^2(4\xi_1+\phi_1)r^{2\xi_1\phi_1-4}=0
\end{equation}
In order for this equation to be satisfied, we must assume that
\begin{equation} \label{x1}
    \xi_1 = \frac{2}{\phi_1}.
\end{equation}
Additionally, we need to fix the charge as follows:
\begin{equation} \label{car}
    q^2 = 2 (z-1) (z+2) l^{-2 z} e^{-\frac{4 \phi_0}{\phi_1}}
\end{equation}
Hence, the electromagnetic field introduced in the action (\ref{action1}) is, in fact, just an auxiliary field. A similar conclusion was obtained in Refs. \cite{Lim:2019hci,Tarrio:2011de}. Finally, the ultimate solution obtained after substituting (\ref{xo}), (\ref{cos}), (\ref{x1}),  and (\ref{car}) into (\ref{ff}) is given by
\begin{equation}
    f(r)= 1 -\frac{m}{r^{z+2}}.
\end{equation}
where the constant $m$ is related to mass, as we will see later.

Considering all of the aforementioned considerations, we can conclude that the solution is characterized by only three free parameters: $m$, $z$, and $\phi_0$.
Summarizing, the solution we found to action (\ref{action1}) is
\begin{equation}\label{resumo1}
    ds^2 =  - \frac{r^{2z}}{l^{2z}} \bigg[ 1 -\frac{m}{r^{z+2}} \bigg] dt^2 + \frac{dr^2}{\frac{r^2}{l^2}\bigg[ 1 -\frac{m}{r^{z+2}} \bigg]} +r^2d\varphi^2 + \frac{r^2}{l^2} dy^2. 
\end{equation}
Note that in the limit $z=1$ and assuming that $l^2=\alpha^{-2}$ 
and $m= \frac{b}{\alpha^{-3}}$, we then recover the well-known solution by Lemos \cite{Lemos:1994xp}.

\subsection{Lifshitz charged black string}

\begin{figure*}
 \includegraphics[height=4.3cm]{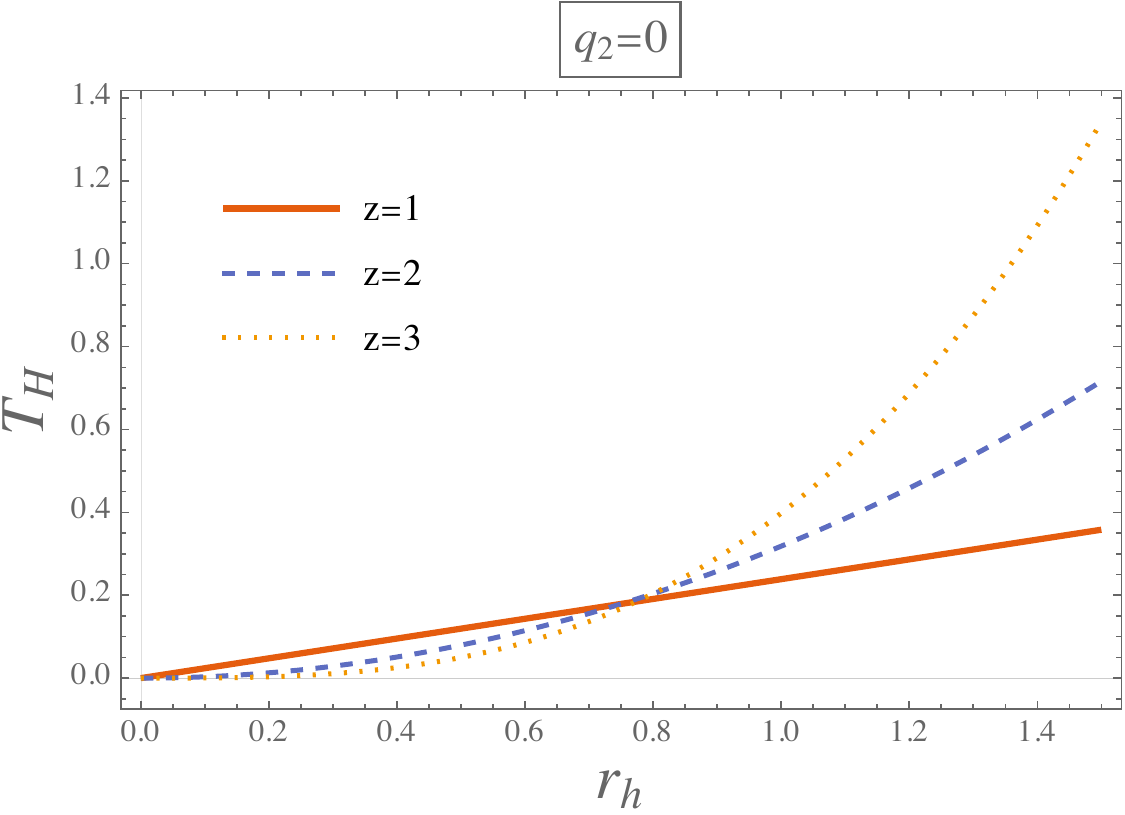}\quad
  \includegraphics[height=4.3cm]{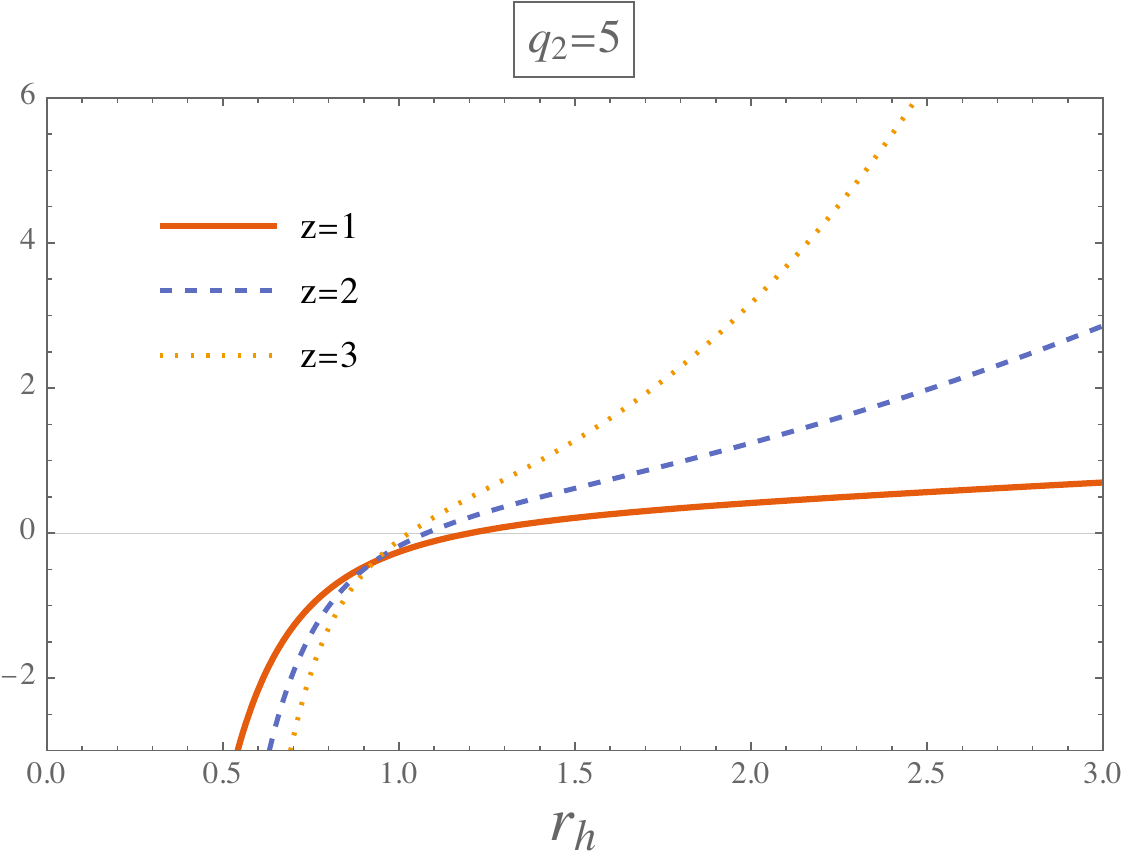}\quad
 \includegraphics[height=4.3cm]{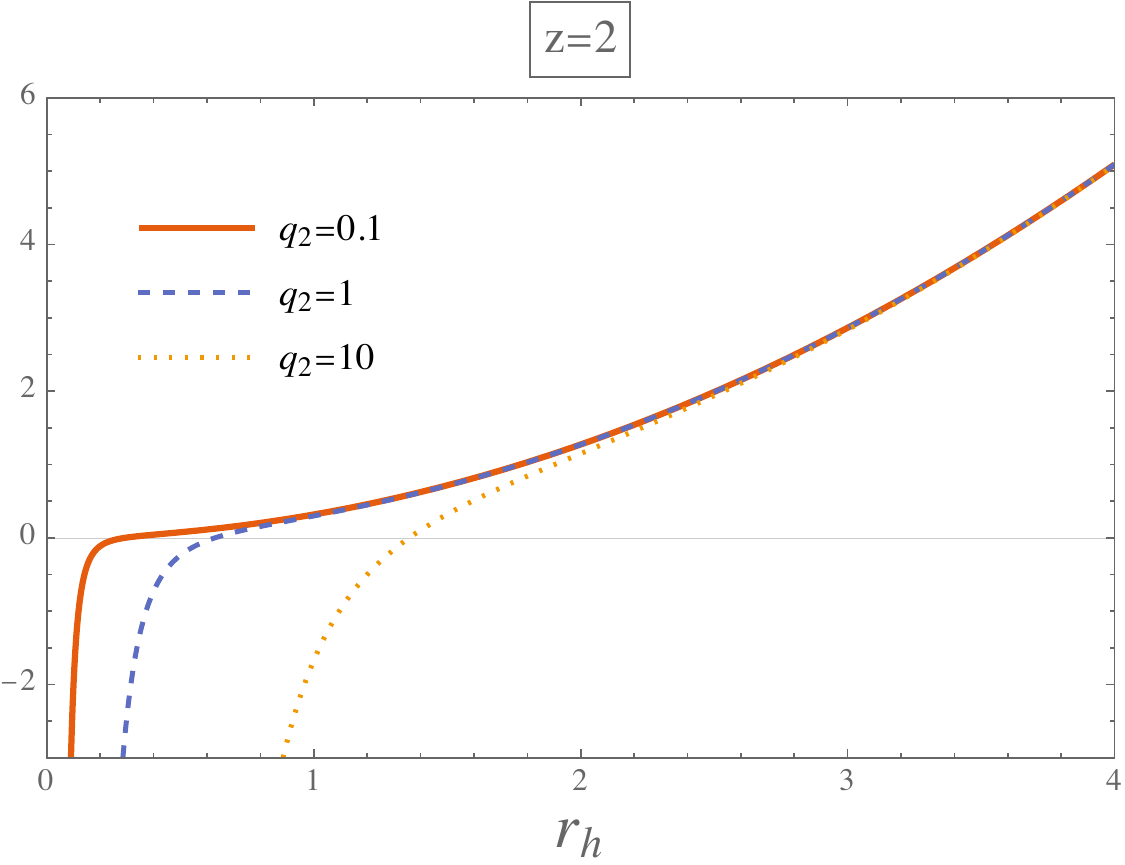}\quad
\caption{Black string's Hawking temperature. We have considered zero electric charge and the first three values of $z$ (left panel), $q_2=0.5$ and the first three values of $z$ (middle panel) and $z=2$ and three values of charge (right panel). We set $l=1$.}  \label{fig1}
\end{figure*}

As seen above, the gauge charge ends up not affecting the charge of the black string, since the electric charge is fixed by the geometry (\ref{car}). However, it is possible to incorporate a contribution from the gauge field into the solution. In this case, one just needs to introduce new gauge fields in (\ref{action1}) that are coupled to the dilaton field, i.e.,  $-\frac{1}{4}F_{i}^2 e^{-2\xi_i\phi}$ ($i=2,3,...$). For simplicity, let's consider only one additional gauge field. In this case, the new electric field can be given by
\begin{equation}\label{camp22}
F_{rt,2} = \frac{q_2 e^{2\xi_2 \phi}}{r^{3-z}},
\end{equation}
where the total charge $U(1)$ is given by 
\begin{equation}\label{carga2}
    Q_2 = \frac{1}{16\pi G} \int e^{-2\xi_2\phi}\star F_{2}.
\end{equation}

It is straightforward to show that the addition of the charge $q_2$ only change Eq. (\ref{ff}) and Eq. (\ref{eq}), as, respectively
\begin{widetext}
\begin{eqnarray}\label{ff2}
   f(r) &=& -\frac{m}{r^{z+2}} +\frac{q^2 e^{2 \xi_1 \phi_0} (\xi_1 \phi_1-1) l^{2 z} r^{2 \xi_1 \phi_1-4}}{4 z (2 \xi_1 \phi_1+z-2)}+\frac{q_2^2 e^{2 \xi_2 \phi_0} (\xi_2 \phi_1-1) l^{2 z} r^{2 \xi_2 \phi_1-4}}{4 z (2 \xi_2 \phi_1+z-2)} -\frac{\Lambda_0  l^2 e^{-2 \xi_0 \phi_0} (\xi_0 \phi_1+1) r^{-2 \xi_0 \phi_1}}{z (-2 \xi_0 \phi_1+z+2)}
\end{eqnarray}
\end{widetext}
and
\begin{eqnarray}\label{eq22}
    \nonumber &&4(1+z)(2+z)\phi_1 - e^{2\xi_1\phi_0}l^{2z}q^2(4\xi_1+\phi_1)r^{2\xi_1\phi_1-4}\\
    &&- e^{2\xi_2\phi_0}l^{2z}q_2^2(4\xi_2+\phi_1)r^{2\xi_2\phi_1-4}=0.
\end{eqnarray}
So that the Eq. (\ref{eq22}) above continues to be satisfied for (\ref{x1}) and (\ref{car}) if
\begin{equation} \label{x2}
    \xi_2 = - \frac{\phi_1}{4}.
\end{equation}

Finally, if we substitute (\ref{x1}), (\ref{car}), and (\ref{x2}) into Eq. (\ref{ff2}), we find the solution for the asymptotically Lifshitz charged black string given by
\begin{equation}\label{resumo1}
    ds^2 =  - \frac{r^{2z}}{l^{2z}} f(r) dt^2 + \frac{dr^2}{\frac{r^2}{l^2}f(r)} +r^2d\varphi^2 + \frac{r^2}{l^2} dy^2, 
\end{equation}
where $f (r)$ is given by
\begin{equation}
    f(r)=1 -\frac{m}{r^{z+2}} + \frac{q_2^2 l^{2 z} e^{-\sqrt{z-1} \phi_0}}{4 z r^{2(z+1)}}
\end{equation}

Now, let us analyze some geometric properties of the solution (\ref{resumo1}). First, we can determine the position of the horizon $r_h$ by setting $f=0$. This results in the following algebraic equation for $r_h$:
\begin{align}\label{hor} 
   1 -\frac{m}{r_h^{z+2}} + \frac{q_2^2 l^{2 z} e^{-\sqrt{z-1} \phi_0}}{4 z r_h^{2(z+1)}}=0.
\end{align}
In this case, we assume that $r_h$ represents the largest positive real root of the equation $f=0$. Unfortunately, an exact solution for $r_h=r_h(m, q_2, \phi_0)$ is not readily available. However, as noted by ref.\cite{Tarrio:2011de}, the mass parameter $m$ is not a fundamental parameter of the theory. Motivated by this, we can solve Eq. (\ref{hor}) with respect to the mass, yielding
\begin{align} \label{massa}
  m=\frac{q_2^2 l^{2 z} r_h^{-z} e^{-\sqrt{z-1} \phi_0 }}{4 z}+r_h^{z+2}.
\end{align}
Additionally, we can compute the Kretschmann scalar near the origin, with the leading term being proportional to $K \sim r^{4(1+z)}$. This indicates that our solution is singular at the origin.

\section{Thermodynamics} \label{3}
Let us now investigate the thermodynamics of the charged black string in Lifshitz spacetime. It is crucial to study the thermodynamic properties of the solution (\ref{resumo1}) and verify the validity of the first law. But first, we set $\phi_0=0$, as this choice does not affect any subsequent thermodynamic analysis since it only pertains to a phase. To proceed, we must formally define the relationship between the parameter $m$ and the mass of the black string. First, let us explicitly define the charge from Eq. (\ref{carga2}) as

\begin{equation} \label{car}
Q = \frac{q_2 l^{z-2}}{8 G_N}.
\end{equation}

Additionally, the potential associated with this charge in the thermodynamic relations, measured at infinity with respect to the horizon, is defined by:

\begin{equation}
\Phi(r) = A_{\mu} \chi^{\mu} \big|_{r \rightarrow \infty} - A_{\mu} \chi^{\mu} \big|_{r = r_h},
\end{equation}	
where $\chi=\partial_t$ is the null generator of the horizon. Considering Eq. (\ref{camp22}) and the equation above, we find that:
\begin{equation} \label{pot}
\Phi(r) = - \frac{q_2 }{z} (r^{-z} - r_h^{-z}).
\end{equation}
 
Using the modified Brown and York formalism \cite{york}, we can calculate the mass of the solution as:
\begin{equation}\label{mass}
    M = \frac{m l^{-z-2}}{4 G_N }.
\end{equation} 

\begin{figure*}
 \includegraphics[height=4.3cm]{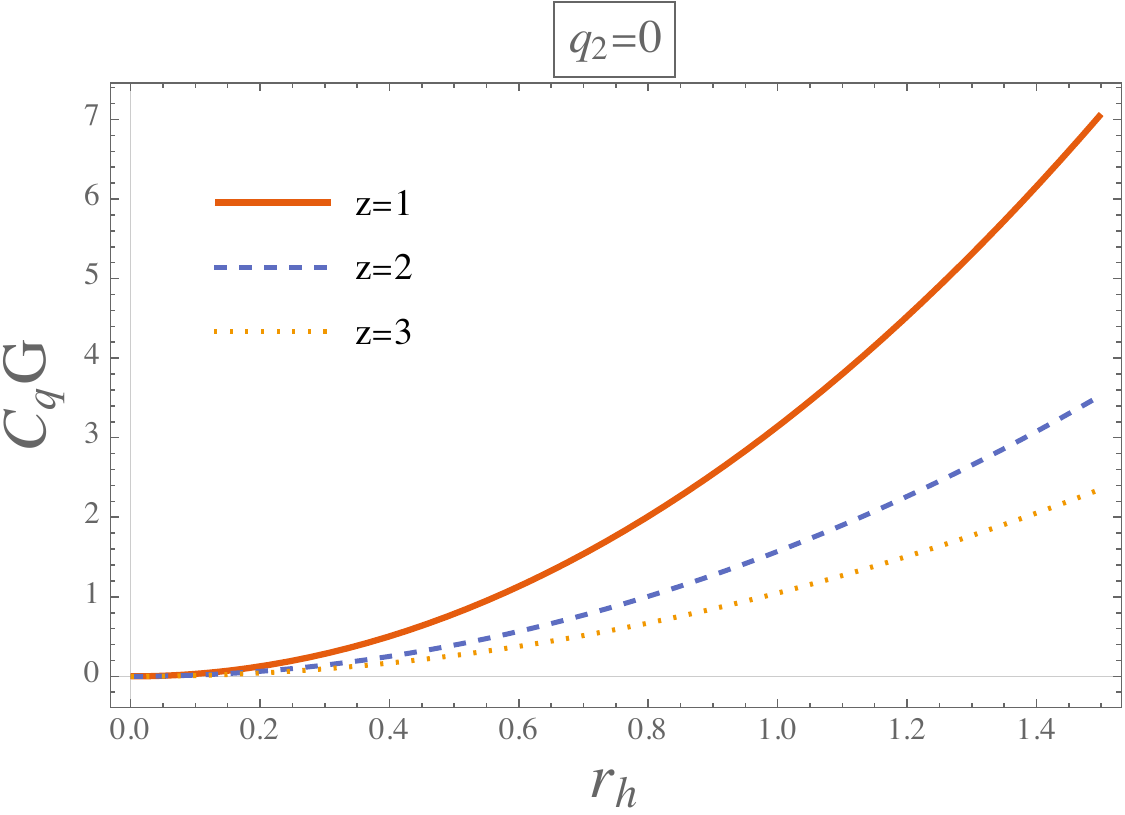}\quad
  \includegraphics[height=4.3cm]{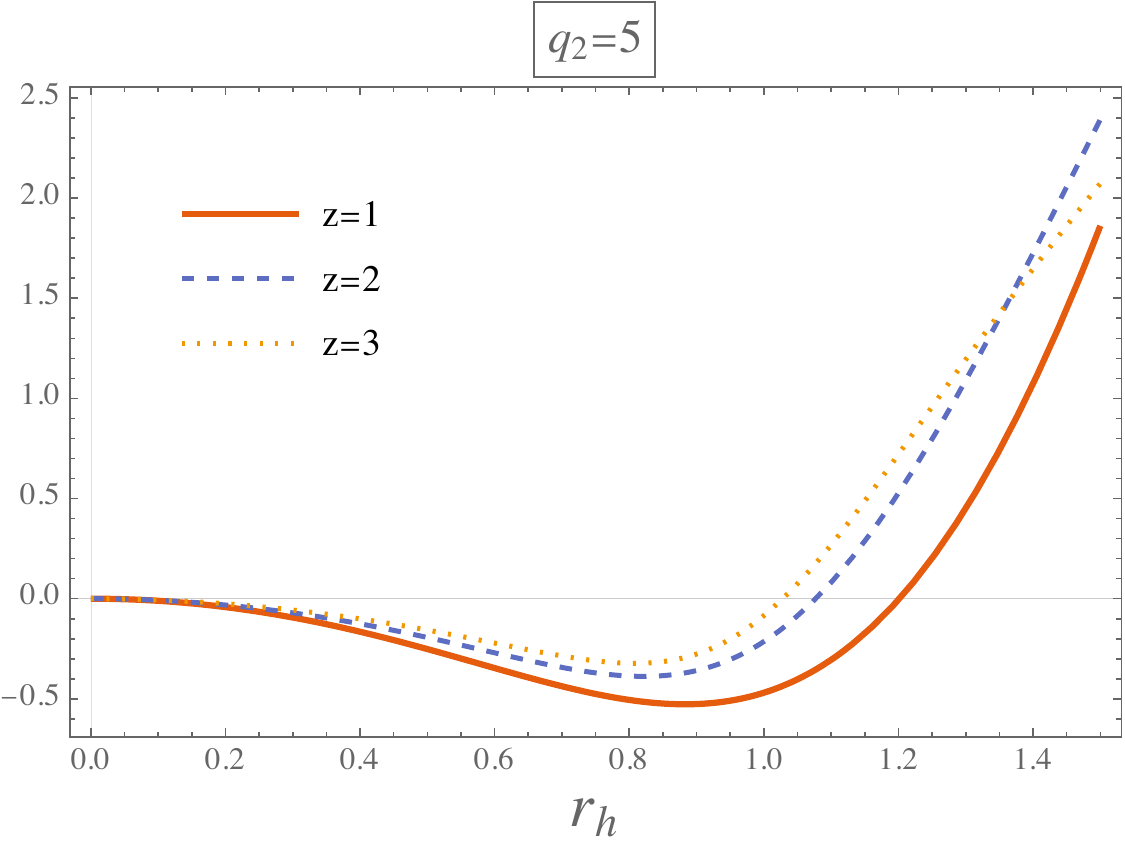}\quad
 \includegraphics[height=4.3cm]{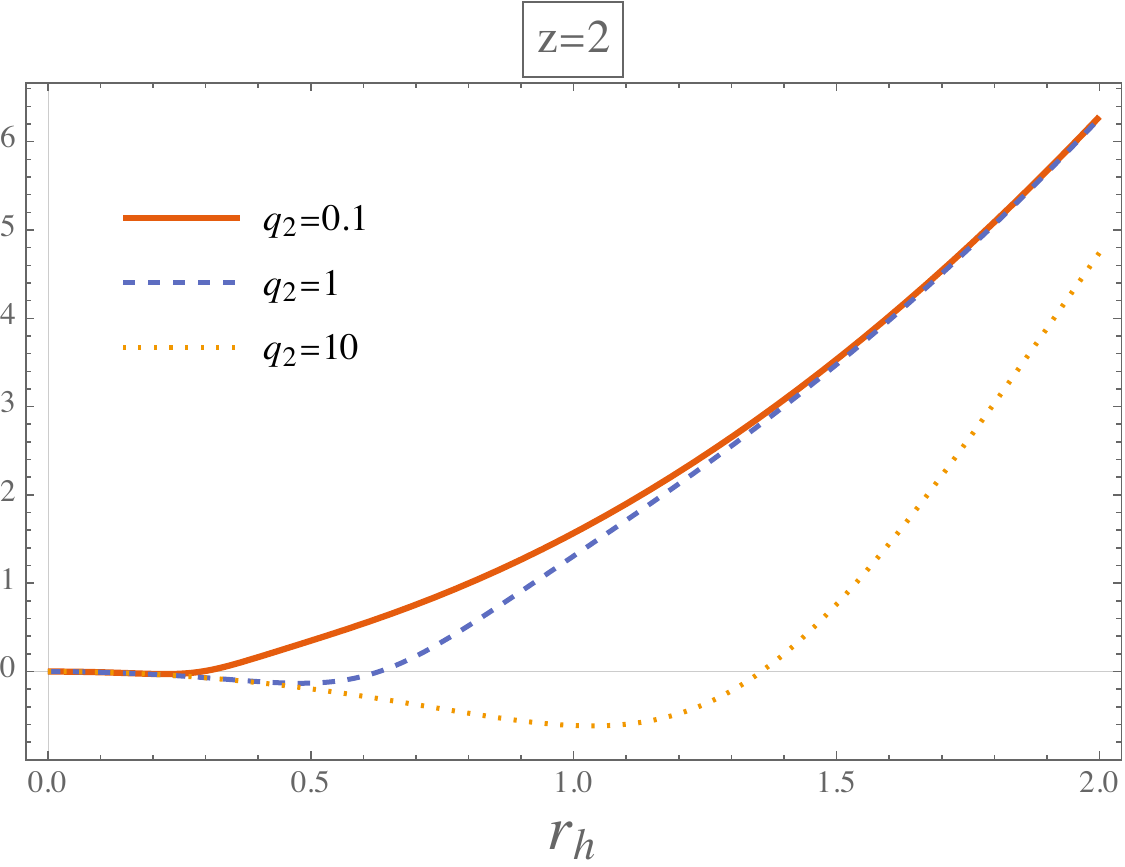}\quad
\caption{Heat capacity. We adopt the same Hawking temperature choices in Fig. (\ref{fig1}). We set $l=1$.}  \label{dia}
\end{figure*}

The black string possesses a Hawking temperature $T_H=\kappa/2\pi$, which can be derived from the surface gravity provided by
\begin{equation}
    \kappa=- \frac{1}{2}(D_{\mu}\chi_{\nu})(D^{\mu}\chi^{\nu})
\end{equation}
where $\chi=\partial_t$ is the Killing vector. Hence, the computed Hawking temperature can be written as
\begin{eqnarray}\label{temp}
    T_H=\frac{l^{-z-1} r_h^{-z} \left(4 (z+2) r_h^{2 z}-\frac{q_2^2 l^{2 z} }{r_h^2}\right)}{16 \pi }.
\end{eqnarray}
From the above expression, we can calculate the charge value at which we reach the extremal limit ($T=0$). This occurs when
\begin{equation}\label{extremalq}
    q_{ext} = 2l^{-z} r_{ext}^{z+1}  \sqrt{(z+2)}
\end{equation}
where $r_{ext}$ is the position of the horizon at extremality, defined by the conditions $f(r_{ext})=f'(r_{ext})=0$. With the expression from (\ref{extremalq}), we can use Eq. (\ref{massa}) to obtain the extremal mass given by
\begin{equation}
    m_{ext} = \frac{2(1+z)}{z}r_{ext}^{2+z}. 
\end{equation}

The behaviour of the Hawking temperature is depicted in Fig. (\ref{fig1}) for three values of the electric charge and the Lifshitz exponent $z$, in order to investigate the influence of the charge and the Lifshitz scaling in the black string's Hawking temperature. In the left panel of Fig. (\ref{fig1}) we have considered $q_2=0$ in order to study the pure influence of the anisotropic scaling. As we can see, when $z=1$ we recover the linear behaviour of the black string's Hawking temperature, as expected. By increasing the value of $z$ the behaviour of the Hawking temperature changes by a simple modification in the power of $r_h$. In the middle and right panels of Fig. (\ref{fig1}) we have considered non-vanishing $q_2$. Notice that the non-vanishing electric charge forces the black string to exhibit a phase transition of zeroth order in which the temperature vanishes and the black string evaporation halts at finite horizon radii. Therefore, remnant masses appear as a result of such phase transitions.

Furthermore, we can use the temperature expression (\ref{temp}) to find the critical points in the canonical ensemble. These points indicate first-order transitions in the system and can be obtained by solving the system of equations generated by the conditions $\frac{\partial T}{\partial r_h}=\frac{\partial ^2 T}{\partial r_h^2}=0$. Directly, we can show that there are no solutions to these conditions; therefore, there are no first-order transitions for  asymptotically Lifshitz charged black strings.

Assuming that the black string solution (\ref{resumo1}) obeys the area law, we have that the Bekenstein-Hawking entropy is given by
\begin{equation} \label{entropia}
S = \frac{\pi r_h^2}{2 G l}.
\end{equation}

Given these thermodynamic quantities, it is straightforward to demonstrate that the first law of thermodynamics holds, namely,
\begin{equation}
    dM = T dS + \Phi(\infty)dQ,
\end{equation}
where $\Phi(\infty) = \frac{q_2 }{z} r_h^{-z} $

To conclude this part of the thermodynamic analysis, let us calculate the heat capacity at constant charge $C_q$ of the charged black string, corresponding to the canonical ensemble, to determine its thermodynamic stability. We know that this stability can be associated with microscopic fluctuations of the system. To analyze this, we need to examine the interval of positivity of the $C_q$, namely,
\begin{eqnarray}
    C_q = T \bigg(\frac{\partial S}{\partial T} \bigg)_{q}\geq0
\end{eqnarray}
Substituting Eqs. (\ref{temp}) and (\ref{entropia}) in the above expression considering $q_2$ fixed, we obtain the thermal capacity of the asymptotically Lifshitz charged black string  given by
\begin{equation}\label{cq}
   C_{q} = \frac{\pi  r_h^2 \left(4 (z+2) r_h^{2 z+2} -q_2^2 l^{2 z}\right)}{G l (z+2) \left(q_2^2 l^{2 z}+4 z r_h^{2 z+2} \right)}
\end{equation}
With the above expression, we can show that thermodynamic stability, i.e., $ C_{q}\geq0$ is achieved when $q_2\leq 
q_{ext}$ (\ref{extremalq}). Further, the heat capacity (\ref{cq}) does not exhibit singular points. Thus, we indeed do not have first-order transitions, as can be verified in the Fig. (\ref{dia}).

\begin{figure}
 \includegraphics[height=4.5cm]{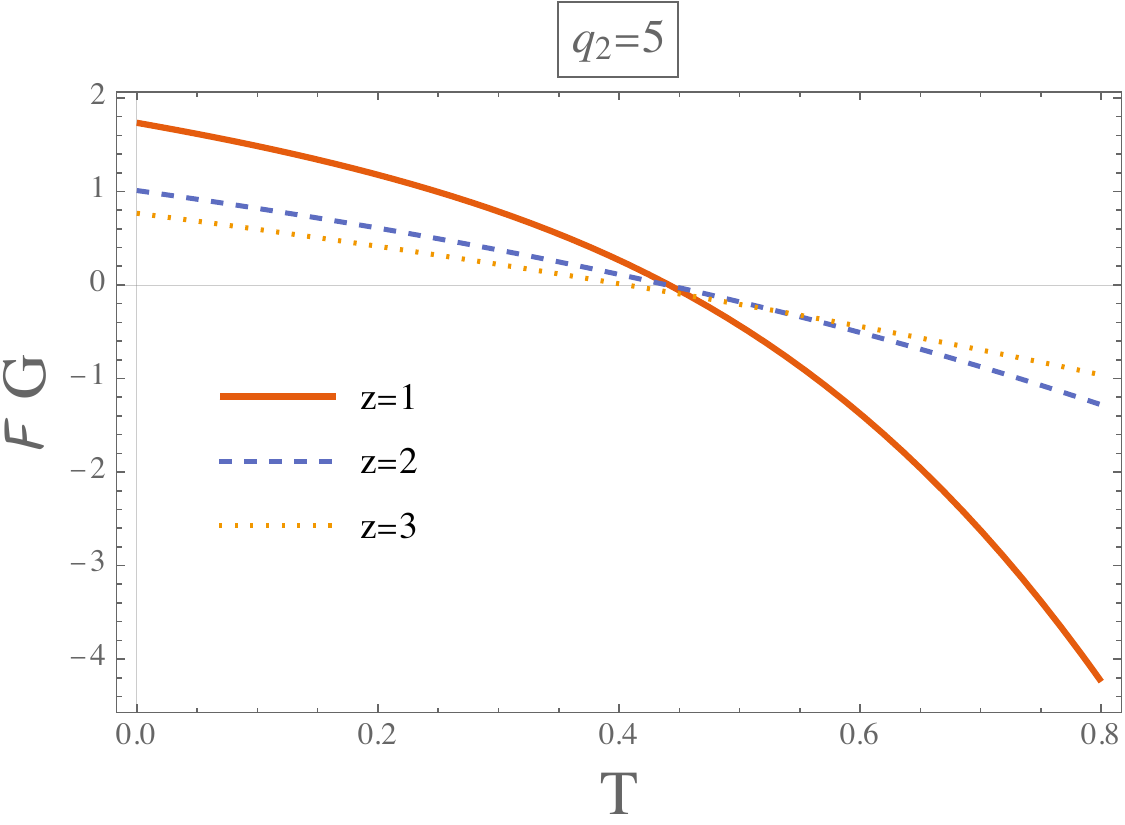}
\caption{ Helmholtz free energy. We set $l=1$.}  \label{helm}
\end{figure}

For completeness, let us calculate the free energy associated with the canonical ensemble, the Helmholtz free energy. This quantity is given by 
\begin{align}
    \mathcal{F}& = \Delta M - T S \\ \nonumber
    &=\frac{ [(2+z)q_2^2l^{2z}-4z^2 r_h^{2(1+z)}]}{32 z G l^{2+z}r_h^z}.
\end{align}
Indeed, as shown in the Fig. (\ref{helm}), we can observe the absence of the swallowtail structure, which is characteristic of first-order transitions

\section{Final remarks}\label{con}

We have shown that it is possible to generate asymptotically Lifshitz black string solutions in the Einstein-Maxwell-Dilaton theory with a cosmological constant. Furthermore, we demonstrated that it is necessary to assume at least two gauge fields to electrically charge the black string. 

The geometric properties of the new solution found depend on the value of the Lifshitz exponent $z$, which must be $z\geq1$ to ensure that the dilaton field is real. However, this field still diverges at the boundary. We found that our solution, which is singular at the origin, has possible horizons that depend on the value of $z$. The new solution recovers Lemos's black string solution when $z=1$. On the other hand, for $z\neq1$, we find new modifications in the horizons and thermodynamic quantities.

To conclude the paper, we investigated the thermodynamic properties of the asymptotically Lifshitz charged black string solution. We calculated the associated Hawking temperature and showed that it increases with $r_h$ as a power of $z$ in the chargeless case. On the other hand, when we consider the effects of electric charge on the black string, we find a zero-order phase transition, resulting in the formation of remnant masses. Additionally, we showed that the solution (\ref{resumo1}) maintains the validity of the first law of thermodynamics.

Finally, we explored the thermodynamic stability of the solution (\ref{resumo1}) with microscopic fluctuations in the canonical ensemble. We calculated the heat capacity of the charged black string and showed that it can be positive, and thus thermodynamically stable, if $q_2\leq 
q_{ext}$, where $q_{ext}$ is the extremal charge (\ref{extremalq}) responsible for $T=0$. Additionally, we did not find any first-order transitions in our solution.

\section*{Acknowledgments}
\hspace{0.5cm} The authors thank the Funda\c{c}\~{a}o Cearense de Apoio ao Desenvolvimento Cient\'{i}fico e Tecnol\'{o}gico (FUNCAP), the Coordena\c{c}\~{a}o de Aperfei\c{c}oamento de Pessoal de N\'{i}vel Superior (CAPES), and the Conselho Nacional de Desenvolvimento Cient\'{i}fico e Tecnol\'{o}gico (CNPq), Grant no. 200879/2022-7 (RVM). 




\end{document}